\documentclass[twocolumn,aps,preprintnumbers,nofootinbib,floatfix,superscriptaddress]{revtex4-1}
\usepackage{amsmath}
\usepackage{graphicx}
\usepackage[normalem]{ulem}
\usepackage{dcolumn}
\usepackage{epsfig}
\usepackage{bm}
\usepackage{array}
\usepackage[english]{babel}
\usepackage{hyperref}
\usepackage{algorithm,algorithmic}
\usepackage{lipsum} 
\usepackage{tikz}
\usetikzlibrary{calc,arrows,chains,matrix,positioning,scopes,external}
\hypersetup{
colorlinks=true,
citecolor=blue,
linkcolor=red,
urlcolor=blue
,pdfmenubar=true
}

\usepackage{amsfonts}
\usepackage{amssymb}
\usepackage{xcolor}
\usepackage{bbm}
\usepackage{calrsfs}
\usepackage{dutchcal}
\usepackage{amsthm}

\newcommand{\bra}[1]{\ensuremath{\langle#1|}}
\newcommand{\ket}[1]{\ensuremath{|#1\rangle}}

\newcommand{\be}{\begin{equation}}
\newcommand{\ee}{\end{equation}}

\newcommand{\beq}{\begin{eqnarray}}
\newcommand{\eeq}{\end{eqnarray}}

\begin{document}

\title{Fundamental bounds for parameter estimation with few measurements}

\author{Valentin~Gebhart}
\affiliation{QSTAR, INO-CNR and LENS, Largo Enrico Fermi 2, 50125 Firenze, Italy}

\author{Manuel Gessner}
\affiliation{Departament de Física Teòrica, IFIC, Universitat de València,
CSIC, C/ Dr. Moliner 50, 46100 Burjassot (València), Spain}

\author{Augusto~Smerzi}
\affiliation{QSTAR, INO-CNR and LENS, Largo Enrico Fermi 2, 50125 Firenze, Italy}


%
\begin{abstract}

Bounding the optimal precision in parameter estimation tasks is of central importance for technological applications.
In the regime of a small number of measurements, or that of low signal-to-noise ratios, the meaning of common frequentist bounds such as the Cramér-Rao bound (CRB) become questionable.
Here, we discuss different linear (Barankin-like) conditions that can be imposed on estimators and analyze when these conditions admit an optimal estimator with finite variance, for any number of measurement repetitions. 
We show that, if the number of imposed conditions is larger than the number of measurement outcomes, there generally does not exist a corresponding estimator with finite variance. 
We analyze this result from different viewpoints and examples and elaborate on connections to the shot-noise limit and the Kitaev phase estimation algorithm.
We then derive an extended Cramér-Rao bound that is compatible with a finite variance in situations where the Barankin bound is undefined. 
Finally, we show an exemplary numerical confrontation between frequentist and Bayesian approaches to parameter estimation.

\end{abstract}

\maketitle

\section{Introduction}\label{sec:introduction}

The theory of parameter estimation treats the question of how to infer parameters from statistical data and thus represents the central piece of inferential statistics~\cite{lehmann2006}, crucial to all scientific fields like physics, engineering, medicine etc. 
Two key concepts of the frequentist approach to parameter estimation are the Fisher information~\cite{fisher1922} and the Cramér--Rao bound (CRB)~\cite{cramer1946,rao1945} that impose fundamental limitations on how precisely unknown parameters can be estimated.
Furthermore, these concepts, if applied to data stemming from measurements of quantum systems, comprise the basis for the fields of quantum metrology and quantum sensing~\cite{helstrom1976,holevo1982,paris2009,giovannetti2011,toth2014,pezze2018}.

The CRB demands certain conditions for an optimal estimator of the unknown parameter and provides a precision bound that becomes attainable only in the limit of many repetitions of a measurement, or in the limit of a large signal-to-noise ratio.
These limitations of the CRB have been addressed in a series of subsequent studies that analyze optimal parameter estimation strategies for different (stronger) conditions imposed on the estimator, and that are applicable for any number of measurement repetitions~\cite{bhattacharyya1946,barankin1949,hammersley1950,chapman1951,abel1993}, extendable also to the quantum domain~\cite{gessner2023}.
In particular, these different bounds have been compared for few repetitions of the measurement, exhibiting a threshold behaviour~\cite{mcaulay1969,mcaulay1971,knockaert1997,renaux2007,chaumette2008}. 

In this article, we analyze the existence of optimal estimators that satisfy different linear conditions of unbiasedness, in the regime of a small number of measurements, where each measurement admits a finite set of possible measurement outcomes.
We derive conditions under which such estimators do exist or not and discuss several examples, among which are the locally unbiased estimators considered by Barankin~\cite{barankin1949}.
Applying these results to different examples in quantum metrology, we compare the conditions to the Kitaev phase estimation algorithm and to the shot-noise limit.
Furthermore, after discussing several properties of the Barankin bound, we combine Barankin's conditions on the estimator with the conditions imposed in the CRB, to obtain and analyze an extended CRB (eCRB) that yields some improvements with respect to the Barankin bound.
Finally, we revisit an exemplary case for which all frequentist bounds exist~\cite{marzetta1993}, and compare the different frequentist bounds with Bayesian bounds on parameter estimation that are usually considered in the threshold regime of a small number of measurements.

\section{Frequentist bounds on parameter estimation}\label{sec:frequentist}

The central objective of parameter estimation is the estimation of an unknown parameter $\theta$ from statistical data that was generated from a probability distribution $p(x|\theta)$ depending on the parameter $\theta$. 
We consider a measurement with $D$ possible outcomes $x=1,\dots,D$, fulfilling $\sum_{x=1}^D p(x|\theta)=1$.
Based on an observation $x$, we employ an estimator $\theta_\mathrm{est}(x)$ to obtain an estimate of the unknown parameter $\theta$. 
The optimal estimator $\theta_\mathrm{est}$ should maximize the precision, i.e., it should minimize the estimation error. 
As a figure of merit, one usually considers the mean squared error of the estimator $\theta_\mathrm{est}$, given by
\begin{equation}\label{eq:var_freq}
    (\Delta \theta_\mathrm{est})^2 = \sum_x p(x|\theta) (\theta_\mathrm{est}(x) -\langle \theta_\mathrm{est}\rangle_\theta)^2
\end{equation}
with $\langle \theta_\mathrm{est}\rangle_\theta = \sum_x p(x|\theta)\theta_\mathrm{est}(x)$. 

Minimizing Eq.~\eqref{eq:var_freq} for a given $\theta$ without further requirements on the estimator $\theta_\mathrm{est}(x)$ would allow for the trivial solution $\theta_\mathrm{est}(x)=\operatorname{const}$ for all $x$, yielding $(\Delta \theta_\mathrm{est})^2=0$. 
Therefore, further requirements on $(\Delta \theta_\mathrm{est})^2$ are needed to sensibly bound the minimal variance $(\Delta \theta_\mathrm{est})^2$. 
In the following, we will consider linear constraints on the estimator $\theta_\mathrm{est}$, i.e., constraints that can be written as
\begin{equation}\label{eq:constraints}
    \lambda = \sum_{x\in X_+} g(x)(\theta_\mathrm{est}(x) -\langle \theta_\mathrm{est}\rangle_\theta),
\end{equation}
where $g(x)$ is a test function corresponding to a bias $\lambda$, and we restricted the sum to the measurement outcomes $x\in X_+$ that occur with nonzero probability $p(x|\theta)>0$. 
As we see later, these constraints may include unbiasedness of the estimator or its derivative for specific parameter values (also called test points). 
To fulfill a set of $n$ constraints corresponding to $(g_k(x),\lambda_k)$ for $k=1,\dots,n$, one can show that the variance is bounded as $(\Delta \theta_\mathrm{est})^2\geq (\Delta \theta_\mathrm{est})_C^2$ with 
\begin{equation}\label{eq:general_bound}
    (\Delta \theta_\mathrm{est})_C^2 = \sup_{\mathbf{a}\in\mathbb{R}^n}\frac{(\mathbf{a}^\top \boldsymbol{\lambda})^2}{\mathbf{a}^\top C \mathbf{a}} = \boldsymbol{\lambda}^\top C^{-1}\boldsymbol{\lambda},
\end{equation}
where $\boldsymbol{\lambda}=(\lambda_1,\dots,\lambda_n)^\top$, and the second equality holds if 
\begin{equation}\label{eq:contraint_matrix}
    C_{kl}=\sum_{x\in X_+}\frac{g_k(x) g_l(x)}{p(x|\theta)}
\end{equation}
is invertible. 
Moreover, the bound identifies the minimum variance of any estimator that satisfies this specific set of constraints. For this reason, a divergence of the bound implies the non-existence of an estimator that satisfies the constraints. For a derivation of Eq.~\eqref{eq:general_bound} see, e.g., Ref.~\cite{gessner2023}.

\subsection{Properties of general frequentist bounds}

Equation~\eqref{eq:general_bound} immediately implies that if there exists an eigenvector $\mathbf{a}_0$ of $C$ with eigenvalue $0$ (meaning that $C$ is not invertible) and such that $\mathbf{a}_0^\top \boldsymbol{\lambda}\neq 0$, the lower bound $(\Delta \theta_\mathrm{est})_C^2$ diverges. 
In other words, the bound diverges if $\Pi_{\ker(C)}\boldsymbol{\lambda}\neq 0$, where $\Pi_{\ker(C)}$ is the projection onto the kernel of $C$, $\ker(C)=\{\mathbf{a}\in\mathbb{R}^n | C\mathbf{a}=0 \}$. 
By defining $\boldsymbol{P}(x)=(g_1(x),\dots,g_n(x))/\sqrt{p(x|\theta)}$ for $x\in X_+$, we can rewrite the matrix $C$ as a finite sum of outer products of the vectors $\boldsymbol{P}(x)$, 
\begin{equation}\label{eq:outer_products}
    C=\sum_{x\in X_+} \boldsymbol{P}(x)\boldsymbol{P}^\top(x). 
\end{equation}
Since each outer product is of rank one, and since $|X_+|\leq D$, we find that $\operatorname{rank}(C)\leq D$. However, $C$ is a $n\times n$ matrix, and, if $D<n$, $C$ possesses at least $n-D$ eigenvectors with corresponding eigenvalue zero. 
Note that $C$ is symmetric and has $n$ real eigenvalues.

We can thus identify two settings that are compatible with a finite bound $(\Delta \theta_\mathrm{est})_C^2$, corresponding to the existence of an estimator (with finite variance) that fulfills the imposed conditions. 
First, if $D\geq n$, the matrix $C$ can be of full rank, meaning that we have a measurement with sufficiently many outcomes ($D$) compared to the number $n$ of imposed conditions on the estimator. 
Second, we can have a non-invertible matrix $C$ (possibly due to $D<n$), but at the same time $\Pi_{\ker(C)}\boldsymbol{\lambda}= 0$, yielding a finite supremum in Eq.~\eqref{eq:general_bound}. 
In this second case, for a given $C$, we emphasize that the set of $\boldsymbol{\lambda}$ that possibly leads to a finite bound corresponds to a $D$-dimensional subspace of $\mathbb{R}^n$, the full $n$-dimensional space of possible choices for $\boldsymbol{\lambda}$. 
Thus, for a randomly chosen $\boldsymbol{\lambda}$, the bound of Eq.~\eqref{eq:general_bound} diverges with certainty if $D<n$.

\subsection{Application: Independent measurements of a single qubit}\label{subsec:qubit}

Let us consider two examples of the situation when the number of imposed conditions is larger than the number of measurement outcomes, i.e., (i) that $D<n$ and the bound of Eq.~\eqref{eq:general_bound} is infinite, or (ii) that $D<n$ but $\Pi_{\ker(C)}\boldsymbol{\lambda}= 0$ and the bound of Eq.~\eqref{eq:general_bound} is finite. 
As conditions for our estimator $\theta_\mathrm{est}$, we take the conditions considered by Barankin~\cite{barankin1949}: 
we impose a number of test points $(\theta_1,\dots,\theta_n)$ at which the estimator should be unbiased. 
This corresponds to the test functions $g_k(x)=p(x|\theta_k)$ and biases $\lambda_k= \theta_k-\theta$ in Eq.~\eqref{eq:constraints}. 
As probability distribution, we consider $m$ independent projective measurements of a single qubit. 
The outcome $\boldsymbol{x}=(x_1,\dots,x_m)\in \{0,1\}^m$ occurs with probability $p_m(\boldsymbol{x}|\theta)=\prod_i p(x_i|\theta)$ with $p(x_i|\theta)=[1+ (-1)^{x_i} r \cos\theta$]/2, where $r$ is the length of the qubit's Bloch vector.
Note that for $m$ independent measurements, the probability of the outcomes are permutationally symmetric, such that two outcomes that are merely permutations of each other contribute the same outer product in Eq.~\eqref{eq:outer_products}. 
Therefore, the rank of the matrix $C$ in Eq.~\eqref{eq:outer_products} is bounded by $D=m+1$, the number of the different outcomes of the corresponding binomial distribution. 

\begin{figure}[t]
		\center
		\includegraphics[width = 0.9\columnwidth]{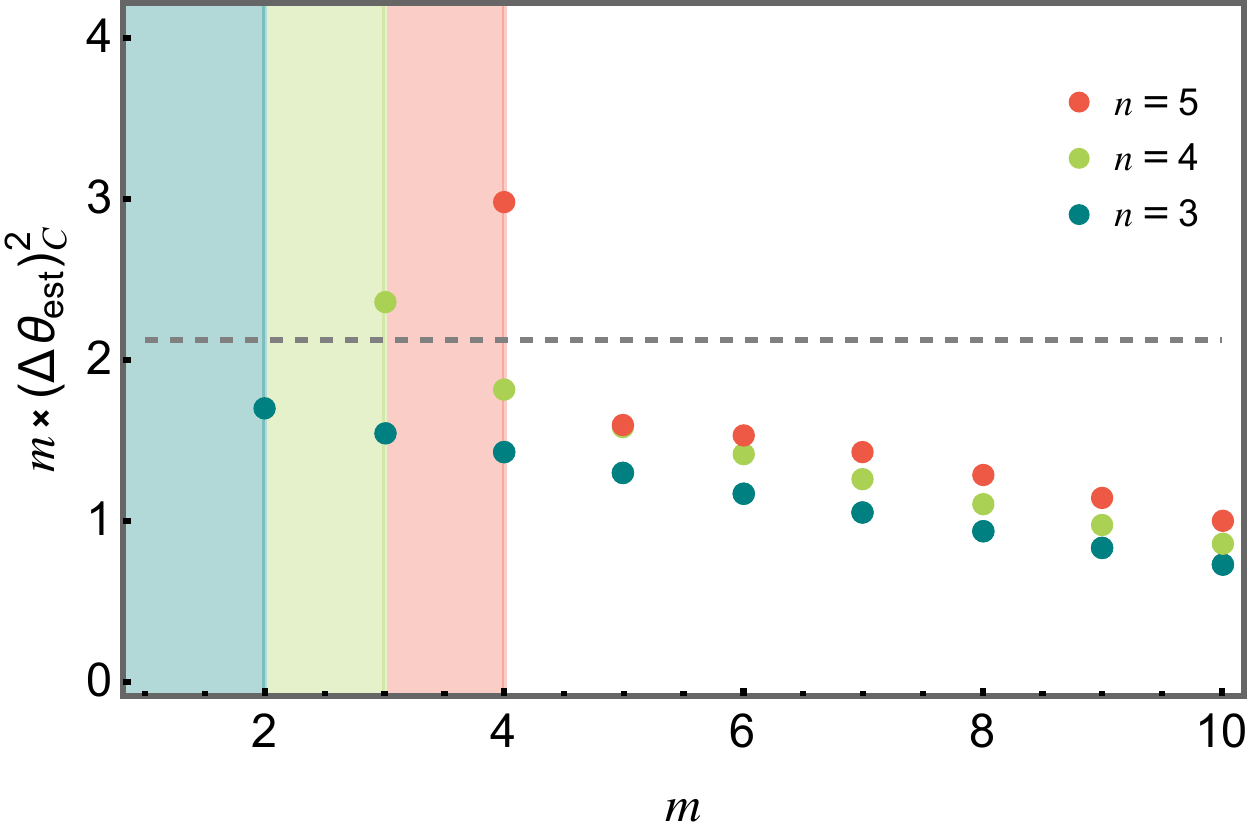} 
		\caption{
            Barankin bound $(\Delta \theta_\mathrm{est})_C^2$, Eq.~\eqref{eq:general_bound}, for $m$ independent measurements of a single qubit with $n$ fixed test points $\theta_k=\theta+(k-1)\pi/6$ for $k=1,\dots,n$ and $n=3$ (blue), $n=4$ (green) and $n=5$ (red) with a true phase $\theta=\pi/4$. The regions for which the Barankin bound diverges because the number of independent outcomes $D$ is smaller than $n$ is indicated by the colored regions. The dashed line is the Cramer-Rao bound (CRB). 
		}
		\label{fig:bar_example}
	\end{figure}

In Fig.~\ref{fig:bar_example}, we show the bound of Eq.~\eqref{eq:general_bound} for the test points $\theta_k=\theta+(k-1)\pi/6$ for $k=1,\dots,n$, with the true phase $\theta=\pi/4$, for $n=3$ (blue), $n=4$ (green), and $n=5$ (red). 
The dashed line corresponds to the CRB, $(\Delta \theta_\mathrm{est})_\mathrm{CRB}^2=(1-r^2\cos^2 \theta)/(m r^2 \sin^2\theta)$. 
For each $n$, the domain where $D=m+1<n$, which generally leads to a divergent bound in Eq.~\eqref{eq:general_bound}, is sketched as a colored region. 
We observe that, as expected, the bound increases when including more test points (i.e., more constraints), and decreases when increasing the number of measurements $m$. 
Furthermore, we note that for any fixed set of test points $(\theta_1,\dots,\theta_n)$, the bound of Eq.~\eqref{eq:general_bound} drops below the CRB (dashed line) for a sufficiently large number of measurements $m$. 
This is due to the fact that the statistical distance between the probability distributions $p(x|\theta_k)$ for the different test points decay exponentially with the number of measurement repetitions $m$, which we further discuss in Sec.~\ref{sec:freq:bar}. Furthermore, as the number of measurements increases, the most stringent condition on the estimator is obtained from infinitesimally separated test points. This indeed corresponds to the Cram\'er-Rao limit, whereas the Barankin-type conditions considered here employ finite spacings between test points.

What about the case where $D<n$ but the bound of Eq.~\eqref{eq:general_bound} is still finite? 
Examples of this case can be constructed by taking an arbitrary estimator $\vartheta_\mathrm{est}(x)$ that leads to a finite variance in Eq.~\eqref{eq:var_freq} (which is guaranteed if the parameter $\theta$ is bounded and the number of measurement outcomes $D$ is finite).
Given the test points $(\theta_1,\dots,\theta_n)$ and the test functions $g_k(x)=p(x|\theta_k)$, we then set the biases as $\lambda_k=\sum_{x} g_k(x)(\vartheta_\mathrm{est}(x) -\langle \vartheta_\mathrm{est}\rangle_\theta)$. 
This means that we use the bias conditions of Eq.~\eqref{eq:constraints}, that restrict ourselves in the optimization, to be automatically fulfilled by the estimator $\vartheta_\mathrm{est}(x)$ we started from, instead of imposing that the estimator must be unbiased at the test points. 
In other words, $\vartheta_\mathrm{est}(x)$ fulfills the requested bias conditions and leads to a finite variance. Therefore, the variance of the best possible estimator $\theta_\mathrm{est}(x)$ fulfilling the conditions, Eq.~\eqref{eq:general_bound}, is bounded by 
\begin{equation}
    (\Delta \theta_\mathrm{est})_C^2 \leq (\Delta \vartheta_\mathrm{est})^2
\end{equation}
and is thus finite. 
Since, in the case of $D>n$, the matrix $C$ in Eq.~\eqref{eq:contraint_matrix} is not invertible, the above analysis shows that here we must have $\Pi_{\ker(C)}\boldsymbol{\lambda}= 0$. 

\subsection{Properties of the Barankin bound}\label{sec:freq:bar}

In the original work of Barankin~\cite{barankin1949}, the minimal variance given specific test points $(\theta_1,\dots,\theta_n)$ is further maximized over the number $n$ of different test points and the value $\theta_k$ of the test points. 
However, if we want to calculate this bound for a specific measurement with $D$ outcomes, we now know that, if $n>D$, the Barankin bound is infinite except for a set of constraints with measure zero. 
Therefore we can conclude that the original Barankin bound that is maximized over the number $n$, and thus in particular includes values $n>D$ for any $D$, diverges. 
We thus conclude that, for any measurement with a finite number of measurement outcomes, there does not exist an estimator that is unbiased on a continuous interval.

Can we further interpret the bound on the number of test points that can be unbiased? If we know that the true phase $\theta$ is part of the interval $\theta\in \left[a,b\right]$ but we have no further information about its precise location, a natural choice for the $n$ test points is to place them equidistantly covering the whole interval, yielding a grid spacing $\Delta \theta = (b-a)/(n-1)$.
Furthermore, for a measurement with $D$ outcomes, we know that the Barankin bound diverges if $D>n$, and thus we find a minimal grid spacing $\Delta \theta$ corresponding to the maximum number of test points $n=D$ such that the Barankin bound is finite, 
\begin{equation}\label{eq:minimal_resolution}
    \Delta \theta = \frac{b-a}{D-1}.
\end{equation}

Let us consider the case where the probabilities correspond to the measurement of $m$ qubits. 
If each qubit is measured in the computational basis, we have $D=2^m$ possible outcomes of the measurement. 
Due to Eq.~\eqref{eq:minimal_resolution}, the minimal grid spacing $\Delta \theta$ scales as $1/2^{m}$, which resembles the precision scaling of the Kitaev phase estimation algorithm \cite{kitaev1995}. 
We believe that this scaling, both in our formalism and in the phase estimation algorithm, can be explained from the fact that, using $m$ bits of information, one can label at most $2^m$ different phases. 
Thus, $m$ bits of information are required to discriminate between $2^m$ different phases, corresponding to a precision scaling of $1/2^m$.
We note that a scaling of $\Delta \theta \sim 1/2^{m}$ does not contradict the Heisenberg limit, $(\Delta \theta)^2 \geq 1/ m^2$, which provides a fundamental lower bound on the precision of standard quantum phase estimation protocols with $m$ qubits. To correctly compare this scaling to the Heisenberg limit, one must count the number of applied phase shifts (i.e., the number of controlled unitary operations in Kitaev's algorithm), instead of the number of qubits~\cite{berry2009}.

If instead we consider $m$-fold repetition of a two-outcome measurement of a single qubit, the outcome probabilities are permutationally symmetric and the corresponding cumulative probability distribution is given by a binomial distribution with $m+1$ different outcomes.
In this case, we find a minimal grid spacing of 
\begin{equation}\label{eq:minimal_resolution_indep}
    \Delta \theta = \frac{b-a}{m}, 
\end{equation}
resembling a Heisenberg-limit scaling of the precision, $(\Delta \theta)^2 \sim 1/m^2$, instead of the shot-noise scaling, $(\Delta \theta)^2 \sim 1/m$, that is the actual precision limit for independent measurements. 
We can explain this apparent contradiction by highlighting that, in the derivation of Eq.~\eqref{eq:minimal_resolution_indep}, we have only used that there are $m+1$ different measurement outcomes. 
In particular, we have not used that these outcomes stem from a binomial distribution that depends on the parameter $\theta$, in which case the probability distributions $p(x|\theta_k)$ corresponding to different grid points $\theta_k$ are highly overlapping. 
This overlap between the different $p(x|\theta_k)$ is the origin of the shot-noise scaling $(\Delta \theta)^2 \sim 1/m$. 
As an example of a measurement with $m+1$ different outcomes that leads to $(\Delta \theta)^2 \sim 1/m^2$, consider the probability distribution $p(x|\theta_k)=\delta_{xk}$, where $\delta$ is the Kronecker symbol. 
Here, the $p(x|\theta_k)$ are orthogonal and, thus, given any outcome we can perfectly discriminate the phase, yielding $(\Delta \theta)^2 \sim 1/m^2$.

Finally, we want to touch upon two further implications about the applicability of Barankin-type bounds. 
If there is an outcome $x_0$ such that $p(x_0|\theta)=1$ (e.g., if we measure a pure quantum state), the optimization of the variance over all estimators that are unbiased at the points $(\theta_1,\dots,\theta_n)$ yields $(\Delta \theta_\mathrm{est})_C^2 = 0$. 
The corresponding optimal estimator fulfills $\theta_\mathrm{est}(x_0)=\theta$ which already yields zero variance in Eq.~\eqref{eq:var_freq}. 
The remaining values of $\theta_\mathrm{est}(x)$ for $x\neq x_0$ are chosen such that the bias conditions on the test points are met (which is always possible because one has an infinite number of degrees of freedom in defining $\theta_\mathrm{est}(x)$ to fulfill a finite number $n$ of conditions). 
Therefore, the significance of achieving a minimal variance under Barankin-like conditions must be questioned. 
For instance, if one has two phases $\theta_1$ and $\theta_2$ which one wants to discriminate, there are well-known bounds on the minimal error probability of the discrimination task,  dependent on the statistical distance between the distributions $p(x|\theta_1)$ and $p(x|\theta_2)$~\cite{nielsen2002}.
In particular, even if there is an $x_0$ such that $p(x_0|\theta_1)=1$, the discrimination is difficult if the two distributions are very similar. 
The apparent conflict with $(\Delta \theta_\mathrm{est})_C^2 = 0$ (if $\theta_1$ is the true phase) is resolved because the estimator optimizing Eq.~\eqref{eq:var_freq} is chosen only to minimize the variance given the true value $\theta_1$, while for the other test point $\theta_2$ the variance might be significantly higher, even though the estimator is unbiased at these points.
Such a highly asymmetric estimator would not be optimal if one considers a balanced discrimination task.

A second and similar characteristic of the Barankin lower bound for fixed test points is the one visible in Fig.~\ref{fig:bar_example} for a large number of measurements $m$. 
Here, for any fixed set of test points $(\theta_1,\dots,\theta_n)$, the Barankin lower bound outpaces the shot-noise-limited CRB for large $m$ (asymptotically, it decays exponentially in $m$). 
This is due to the fact that, for large $m$, the probability distributions corresponding to the different test points become more orthogonal such that one can simply set $\theta_\mathrm{est}(x)=\theta_k$ if $x$ is in the support of $p(x|\theta_k)$. 
Again, it is questionable how valuable such an estimator is.

One way of creating a sensible lower bound on the variance is the one suggested by Barankin, i.e., optimizing the bound over the number $n$ of test points and their values. 
However, as we have seen above, this generally leads to a divergent lower bound if the number of outcomes is finite. 
An intermediate idea is to fix the number of test points $n$ but optimize over their values. 
For instance, in the case of $n=2$, this leads to the Hammersley--Chapman--Robbins bound \cite{hammersley1950,chapman1951,gessner2023}. 
We have analysed this optimization for $n>2$ for repeated measurements of a single qubit, see Sec.~\ref{subsec:qubit}, but encountered numerical problems because, when the test points and their probability distribution become very similar, the matrix C in Eq.~\eqref{eq:contraint_matrix} becomes noninvertible. 
Therefore, even though the optimization over the test points may be finite, a numerical computation is challenging.

\subsection{Extended Cramer--Rao bound}

For the Barankin bound with fixed test points, one requires the estimator $\theta_\mathrm{est}(x)$ to be unbiased at the test points $(\theta_1,\dots,\theta_n)$ (one of which is the true phase $\theta$). 
Instead, to derive the CRB, one only requires that $\partial_\theta \langle \theta_\mathrm{est}\rangle_\theta = 1$, i.e., that the derivative of the mean of the estimator, at the value of the true phase, is one. 
In terms of the test functions of Eq.~\eqref{eq:constraints}, this corresponds to $g(x)=\partial_\theta p(x|\theta)$ and $\lambda=1$. 
This requirement can be directly extended to $n$ different test points $(\theta_1,\dots,\theta_n)$ by requiring the constraints of Eq.~\eqref{eq:constraints} with $g_k(x)=\partial_\theta p(x|\theta)\rvert_{\theta=\theta_k}$ and $\lambda_k=1$ for $k=1,\dots,n$. 
The matrix of Eq.~\eqref{eq:contraint_matrix} corresponding to this extended CRB (eCRB) is given by 
\begin{equation}\label{eq:eCRB}
    (C_\mathrm{eCRB})_{kl}= \sum_{x\in X_+} \frac{\partial_\theta p(x|\theta)\rvert_{\theta=\theta_k} \partial_\theta p(x|\theta)\rvert_{\theta=\theta_l} }{p(x|\theta)}. 
\end{equation}
We note that, despite its apparent similarity, this matrix differs from the Fisher information matrix~\cite{helstrom1976}. The latter describes the case of a probability distribution that depends on $n$ different phases and the derivatives are made with respect to these different phases, while each entry is finally evaluated at the same phases. The quantum bound corresponding to Eq.~\eqref{eq:eCRB} is discussed in Appendix~\ref{app:quantumbound}.

\begin{figure}[t]
		\center
		\includegraphics[width = 0.9\columnwidth]{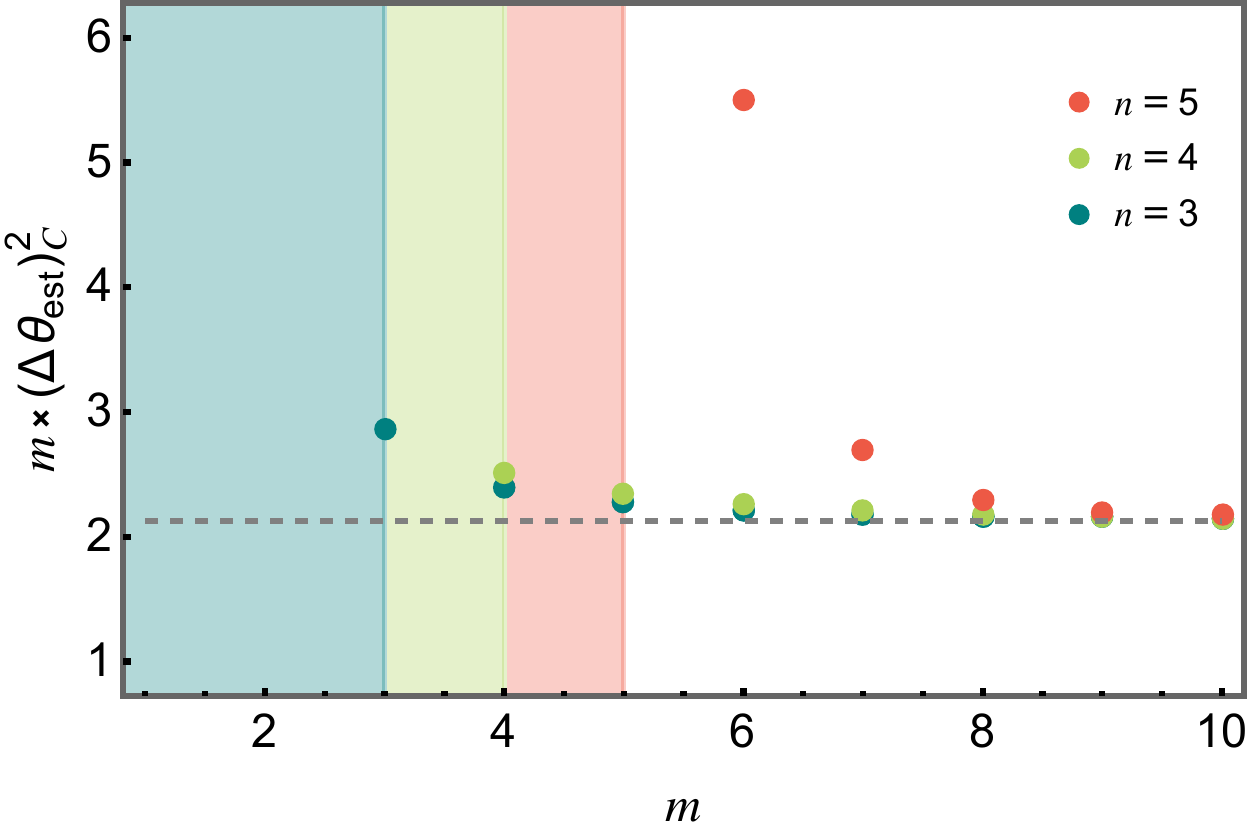} 
		\caption{
		    Extended Cramer-Rao bound (eCRB) $(\Delta \theta_\mathrm{est})_C^2$, Eq.~\eqref{eq:general_bound}, for $m$ independent measurements of a single qubit with $n$ fixed test points $\theta_k=\theta+(k-1)\pi/6$ for $k=1,\dots,n$ and $n=3$ (blue), $n=4$ (green) and $n=5$ (red) with a true phase $\theta=\pi/4$. In contrast to the Barankin bound shown in Fig.~\ref{fig:bar_example}, here we require that $\partial_\theta \langle \theta_\mathrm{est}\rangle_{\theta_k} = 1$ for $k=1,\dots,n$. The regions for which the bound diverges because the number of independent outcomes $D$ is smaller than $n-1$ is indicated by the colored regions. The dashed line is the Cramer-Rao bound (CRB). 
		}
		\label{fig:bar_der}
	\end{figure}

Different to the Barankin bound, the eCRB is always bounded from below by the CRB because it includes the same condition ($\partial_\theta \langle \theta_\mathrm{est}\rangle_\theta = 1$). 
This behaviour can be seen in Fig.~\ref{fig:bar_der}, where we show the eCRB for the same measurement probabilities of $m$ independent measurements of a single qubit and for the same test points as were used in Fig.~\ref{fig:bar_example} for the Barankin bound. 
Therefore, with the eCRB, we have circumvented one of the properties of the Barankin bound, i.e., that, for large $m$, the Barankin bound drops below the CRB because the probability distributions $p(x|\theta_k)$ become orthogonal.
However, also for the eCRB, if the number of imposed conditions is too large, the matrix $C_\mathrm{eCRB}$ is generally not invertible.
Actually, this happens already for $D<n+1$ (requiring one more outcome for a given $n$ with respect to the Barankin bound) because, in the sum of outer products in Eq.~\eqref{eq:outer_products}, the outer products are not linearly independent:
Using that $\sum_x g_k(x)=\sum_x \partial_\theta p(x|\theta)\rvert_{\theta=\theta_k}=0$, one can write, for any outcome $x_0\in X_+$,
\begin{equation}
    \boldsymbol{P}(x_0)\boldsymbol{P}^\top(x_0) =\sum_{x,y\neq x_0} \boldsymbol{P}(x)\boldsymbol{P}^\top(y). 
\end{equation}
Therefore, subtracting the term $\boldsymbol{P}(x_0)\boldsymbol{P}^\top(x_0)$ from Eq.~\eqref{eq:outer_products} does not necessarily  decrease the rank of $C_\mathrm{eCRB}$, which is now a sum of only $D-1$ outer products, and thus $\operatorname{rank}(C_\mathrm{ECRB})\leq D-1$. 

\section{Comparison with Bayesian parameter estimation}

For a small number of measurements, it is generally thought to be more sensible to use Bayesian estimation techniques instead of frequentist ones, see, e.g., Refs.~\cite{hajek2023,li2018} for a discussion. 
Here, one assumes a Bayesian prior distribution $p_\mathrm{prior}(\theta)$, representing the knowledge about the parameter $\theta$ before any measurement. 
The prior is then updated according to Bayes rule, resulting in the Bayesian posterior distribution $p_\mathrm{post}(\theta|x)$ after observing the measurement outcome $x$, given by
\begin{equation}\label{eq:bayes_theorem}
    p_\mathrm{post}(\theta|x) = \frac{p(x|\theta)p_\mathrm{prior}(\theta)}{\int \mathrm{d}\theta p(x|\theta)p_\mathrm{prior}(\theta)}.
\end{equation}
Such a posterior distribution then yields the Bayesian variance $(\Delta \theta)^2_\mathrm{Bayes}(x)$, i.e., the variance of $p_\mathrm{post}(\theta|x)$ for an observed measurement outcome $x$, defined by 
\begin{equation}\label{eq:bay_var}
    (\Delta \theta)^2_\mathrm{Bayes}(x) = \int \mathrm{d}\theta p_\mathrm{post}(\theta|x) (\theta - \langle\theta\rangle_{p_\mathrm{post}(\theta|x)})^2,
\end{equation}
where $\langle\theta\rangle_{p_\mathrm{post}(\theta|x)} = \int \mathrm{d}\theta p_\mathrm{post}(\theta|x) \theta$.

It is widely believed that frequentist and Bayesian approaches cannot be compared due to their conceptual difference of attributing meaning to probability~\cite{hajek2023,li2018}. 
However, we can at least numerically compare the two approaches for an exemplary case. 
To also include the Barankin bound in the comparison, we choose a probability distribution $p(x|\theta)$ that yields a finite Barankin bound, where we intend the original Barankin bound optimized over the number $n$ of test points and their values $(\theta_1,\dots,\theta_n)$, 
\begin{equation}\label{eq:Bar_optimized}
    (\Delta \theta_\mathrm{est})^2_\mathrm{Bar}=\sup_{n\in\mathbb{N},(\theta_1,\dots,\theta_n)} \boldsymbol{\lambda}^\top C^{-1}\boldsymbol{\lambda},  
\end{equation}
see Eq.~\eqref{eq:general_bound} for the definitions of $\boldsymbol{\lambda}$ and $C$. 
We know from Sec.~\ref{sec:frequentist} that if we consider a measurement with an infinite number of outcomes, we may be able to avoid divergences due to matrix rank-deficiency. 

The example we choose is the estimation of the zero-clicks probability of a Poisson distribution, for which the Barankin bound is finite and can be calculated analytically~\cite{marzetta1993}. 
In particular, we consider $m$ independent measurement outcomes $(x_1,\dots,x_m)$, each of which is distributed as
\begin{equation}\label{eq:p_poisson}
    p(x|\theta)=\frac{(-\ln{\theta})^x \theta}{x !}, 
\end{equation}
where $\theta\in[0,1]$ and $x\in\mathbb{N}_0$. 
Here, the Poisson distribution is parametrized by the probability of observing $x=0$, $\theta=p(0|\theta)$.
The Barankin bound of Eq.~\eqref{eq:Bar_optimized} can be computed analytically and is given by~\cite{marzetta1993}
\begin{equation}\label{eq:bar_poisson}
    (\Delta \theta_\mathrm{est})^2_\mathrm{Bar} = \theta^2\left(\theta^{-1/m}-1\right). 
\end{equation}
Furthermore, we compare the frequentist Barankin and the Bayesian variance to the frequentist variance of the widely-used maximum-likelihood estimator (MLE) defined as $\theta_\mathrm{MLE}(x)=\operatorname{argmax}_{\theta\in[0,1]}p(x|\theta)$.
In our case, it is given by 
\begin{equation}\label{eq:mle_poisson}
    (\Delta \theta_\mathrm{est})^2_\mathrm{MLE} = \theta^{m(1-e^{-2/m})} - \theta^{2m(1-e^{-1/m})}.
\end{equation}
Finally, we include the frequentist CRB~\cite{cramer1946,rao1945}
\begin{equation}\label{eq:CRB}
    (\Delta \theta_\mathrm{est})^2_\mathrm{CRB} = \frac{1}{mF(\theta)}=\frac{-\theta^2\ln{\theta}}{m} 
\end{equation}
in our comparison, where $F(\theta)=\sum_x p(x|\theta)\left[\partial_\theta \ln{p(x|\theta)}\right]^2$ is the Fisher information of $p(x|\theta)$ at the parameter $\theta$.

\begin{figure}[t]
		\center
		\includegraphics[width = 0.9\columnwidth]{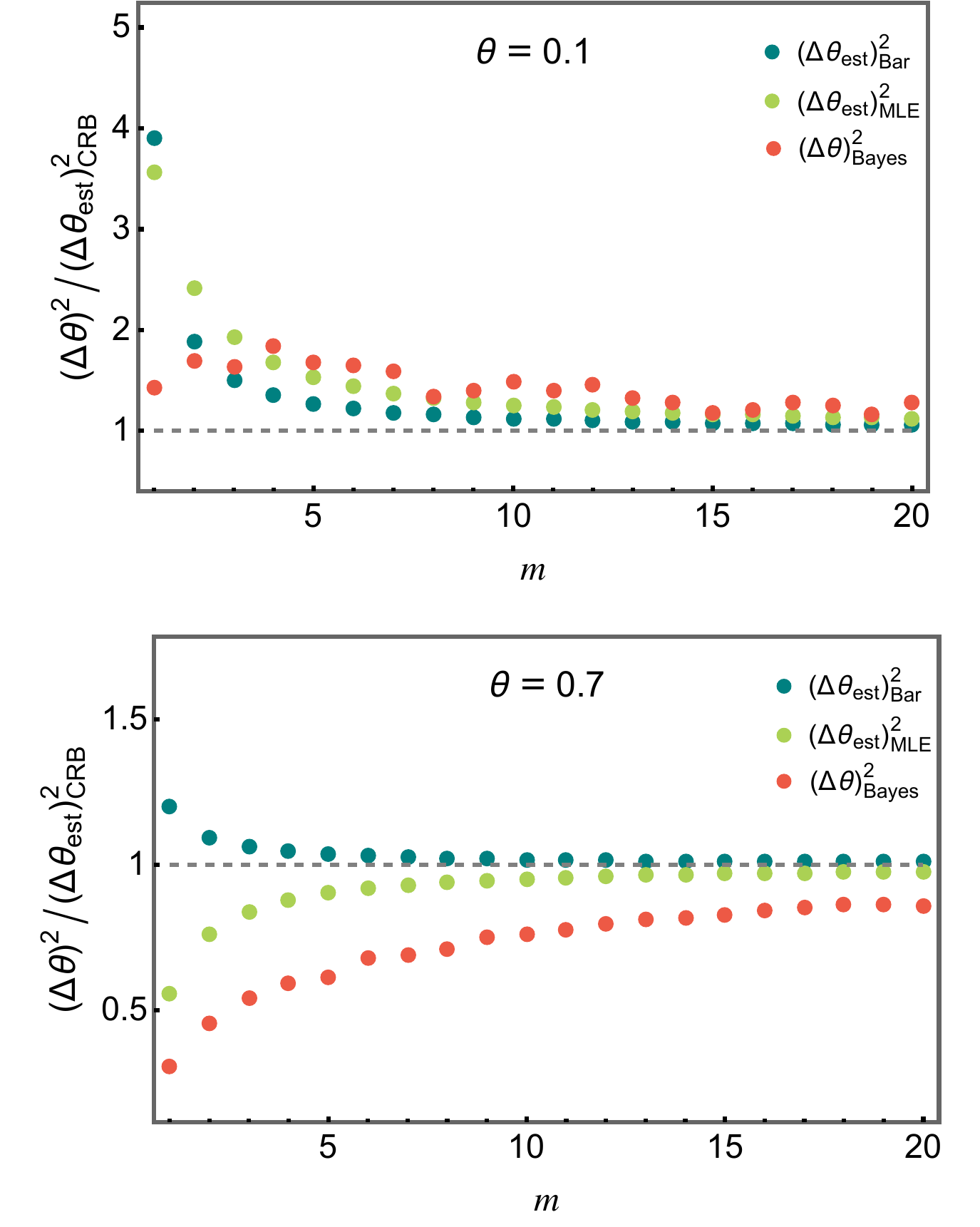} 
		\caption{
		    Comparison of the frequentist Barankin bound, $(\Delta \theta_\mathrm{est})^2_\mathrm{Bar}$, (Eq.~\eqref{eq:bar_poisson}; blue), the frequentist variance of the MLE, $(\Delta \theta_\mathrm{est})^2_\mathrm{MLE}$, (Eq.~\eqref{eq:mle_poisson}; green) and the Bayesian variance, $(\Delta \theta)^2_\mathrm{Bayes}$, (Eq.~\eqref{eq:bay_var}; red), each normalized to the frequentist Cramer-Rao bound $(\Delta \theta_\mathrm{est})^2_\mathrm{CRB}$ (Eq.~\eqref{eq:CRB}, dashed line), as a function of $m$, the number of independent measurements. The underlying probability distribution is Possonian, see Eq.~\eqref{eq:p_poisson}, and the chosen parameter value is $\theta=0.1$ in the upper and $\theta=0.7$ in the lower panel.
		}
		\label{fig:bayesian}
	\end{figure}

In Fig.~\ref{fig:bayesian}, we show the frequentist variances $(\Delta \theta_\mathrm{est})^2_\mathrm{Bar}$ of Eq.~\eqref{eq:bar_poisson} (blue) and $(\Delta \theta_\mathrm{est})^2_\mathrm{MLE}$ of Eq.~\eqref{eq:mle_poisson} (green), normalized by the CRB $(\Delta \theta_\mathrm{est})^2_\mathrm{CRB}$, for two exemplary values of the parameter $\theta$ ($\theta=0.1$ in the upper and $\theta=0.7$ in the lower panel). 
Furthermore, in red, we show the Bayesian variance $(\Delta \theta)^2_\mathrm{Bayes}(x)$ (also normalized by the CRB), averaged over $200$ random outcomes $x$ each of which is generated from the distribution $p(x|\theta)$ in Eq.~\eqref{eq:p_poisson} for the corresponding value of $\theta$.
Note that we use a flat prior in Eq.~\eqref{eq:bayes_theorem}.

First, we can see that all (frequentist or Bayesian) variances approach the CRB for a large number of measurements $m$, as expected. 
Furthermore, the Barankin bound is always larger than the Cramer-Rao bound, as it is an upper bound for the latter. 
Instead, for small $m$, the variance of the MLE is not restricted by the Barankin bound or the CRB, i.e., depending on the value of $\theta$, it may be above the Barankin bound or below the CRB ~\cite{li2018}. 
Whenever the variance of the MLE is below the Barankin bound, it may be of more practical use than the Barankin estimator. 
In particular, for all values of $m$, we checked that the bias of the MLE is much lower than the (square root of the) variance of the MLE (not shown in the figure). 
This means that even though the MLE is biased, the bias is small compared to the statistical fluctuations, and thus the MLE may be useful also in this regime. 
Finally,  
we show that, while, according to the Bernstein–von Mises theorem and under certain regularity conditions for the prior, the  Bayesian and the frequentist inferences coincide asymptotically in the number of measurements, in the case of low signal to noise ratios the two paradigms provide drastically different results. This, however, is also related to the fact that for a small number of measurements the frequentist probabilities are not Gaussian and therefore the variance does not provide the standard deviation probabilistic interpretation. 
We also want to note that, from the quantum viewpoint, the Poisson distribution of Eq.~\eqref{eq:p_poisson} results from the photon-number-resolving measurement of a coherent state $\ket{\alpha}$ with coherent amplitude $\alpha=\sqrt{-\ln{\theta}}$. 
In this case, one can further compare the above bounds to the quantum versions of the bounds~\cite{gessner2023}, see Appendix~\ref{ap:quantum_poisson} for further details.

\section{Conclusions}\label{sec:conclusions}
In this work, we have analyzed the existence of optimal estimators with finite variance for measurements with a finite number of outcomes, if the estimators are restricted to a number of linear Barankin-like conditions.
We found that, if the number of imposed conditions is larger than the number of independent measurement outcomes, such estimators generally do not exist. 
This result can be applied to different bounds in the literature, such as the ones considered by Barankin~\cite{barankin1949}, Bhattacharyya~\cite{bhattacharyya1946}, or Hammersley, Chapman and Robins~\cite{hammersley1950,chapman1951,gessner2023}.
If the measurement probabilities stem from the measurement of several qubits, our bound can be connected to the precision scaling of the Kitaev phase estimation algorithm and, if the measurements are independent, our results can be confronted with the Cramér-Rao bound (CRB). 
Furthermore, we discuss several issues that may arise in the calculation of the Barankin bound and partially address them by introducing an extended CRB. 
Finally, we conduct a numerical comparison between different frequentist bounds on parameter estimation, such as the Barankin bound or the CRB, and the Bayesian approach to parameter estimation. 
To conclude, we notice that the extended bounds discussed in this article have been obtained considering unbiased estimators. We emphasize that any biased estimator would be useful also in the regime of a small number of measurements, whenever the bias is much smaller compared to the variance. Notice that this is in fact the case in the asymptotic regime where the maximum likelihood estimator is, in general, biased for any large but finite number of measurements, being unbiased only asymptotically.

\section*{Acknowledgments}
This work was supported by the European Commission through the H2020 QuantERA ERA-NET Cofund in Quantum Technologies project “MENTA”. This work was funded by MCIN/AEI/10.13039/501100011033 and the European Union 'NextGenerationEU' PRTR fund [RYC2021-031094-I]. This work was funded by the Ministry of Economic Affairs and Digital Transformation of the Spanish Government through the QUANTUM ENIA project call—QUANTUM SPAIN project, by the European Union through the Recovery, Transformation and Resilience Plan—NextGenerationEU within the framework of the Digital Spain 2026 Agenda, and by the CSIC Interdisciplinary Thematic Platform (PTI+) on Quantum Technologies (PTI-QTEP+).

\appendix
\section*{Appendix}

\section{Quantum extended Cram\'er-Rao bound}\label{app:quantumbound}
The classical bound 
\begin{align}
    (\Delta\theta_{\mathrm{est}})^2\geq (\Delta\theta)_{\mathrm{eCRB}},
\end{align}
with 
\begin{align}
    (\Delta\theta)_{\mathrm{eCRB}}=\sup_{\mathbf{a}\in\mathbb{R}^n}\frac{(\mathbf{a}^\top \boldsymbol{\lambda})^2}{\mathbf{a}^\top C_{\mathrm{eCRB}} \mathbf{a}} = \boldsymbol{\lambda}^\top C_{\mathrm{eCRB}}^{-1}\boldsymbol{\lambda}
\end{align}
and
\begin{align}
    (C_\mathrm{eCRB})_{kl}= \sum_{x\in X_+} \frac{\partial_\theta p(x|\theta)\rvert_{\theta=\theta_k} \partial_\theta p(x|\theta)\rvert_{\theta=\theta_l} }{p(x|\theta)},
\end{align}
can be quantized by optimizing over all possible POVMs. This can be done analytically using the method presented in Ref.~\cite{gessner2023}, and yields
\begin{align}
    (\Delta\theta)^2_{Q_\mathrm{eCRB}}&:=\inf_{E_x}(\Delta\theta)^2_{\mathrm{eCRB}}\notag\\&\:=\sup_{\mathbf{a}\in\mathbb{R}^n}\frac{(\mathbf{a}^\top \boldsymbol{\lambda})^2}{\mathbf{a}^\top Q_{\mathrm{eCRB}} \mathbf{a}} = \boldsymbol{\lambda}^\top Q_{\mathrm{eCRB}}^{-1}\boldsymbol{\lambda},
\end{align}
where
\begin{align}
    (Q_{\mathrm{eCRB}})_{kl}=\mathrm{Tr}\left\{\left(\left.\frac{\partial}{\partial\theta}\rho(\theta)\right|_{\theta=\theta_k}\right)\Omega_{\rho(\theta)}\left(\left.\frac{\partial}{\partial\theta}\rho(\theta)\right|_{\theta=\theta_l}\right)\right\}.
\end{align}
Here, $\Omega_{\rho}(X)$ is a superoperator defined as
\begin{align}
    \Omega_{\rho}(X)=\sum_{i,j}\frac{2}{p_i+p_j}|i\rangle\langle i|X|j\rangle\langle j|,
\end{align}
and $\rho=\sum_ip_i|i\rangle\langle i|$ is a full-rank density matrix.

\section{Comparison of (quantum) bounds corresponding to a Poisson distribution}\label{ap:quantum_poisson}
As mentioned in the main text, the Poisson distribution of Eq.~\eqref{eq:p_poisson} can be generated by a number-resolving measurement of a coherent state. In particular, for the coherent state
\begin{equation}\label{eq:coherent}
    \ket{\psi_\theta} = \sum_n \frac{(\sqrt{-\ln{\theta}})^n}{\sqrt{n!}}\ket{n}, 
\end{equation}
the number distribution given by $p(n|\theta) = \left|\left\langle n |\psi_\theta\right\rangle\right|^2$ is given by Eq.~\eqref{eq:p_poisson}. In Ref.~\cite{gessner2023}, the classical bounds (i.e., Barankin, Battacharyya, and Abel bounds) are generalized to the quantum domain. In particular, for a given quantum state $\ket{\psi_\theta}$ that is parametrized by the parameter $\theta$, one minimizes the classical bound over the choice of measurement observable. For $m$ independent measurements of a pure state, all quantum bound $(\Delta \theta_\mathrm{est})_Q^2$ are identical to the quantum CRB, $(\Delta \theta_\mathrm{est})_{Q_\mathrm{CRB}}^2 $, 
\begin{equation}
    (\Delta \theta_\mathrm{est})_Q^2 = (\Delta \theta_\mathrm{est})_{Q_\mathrm{CRB}}^2 =\frac{1}{m F_Q(\ket{\psi_\theta}\bra{\psi_\theta})},
\end{equation}
where $F_Q$ is the quantum Fisher information. For the coherent state of Eq.~\eqref{eq:coherent}, we find \cite{pezze2018}
\begin{align}
    F_Q(\ket{\psi_\theta}\bra{\psi_\theta}) &= 4\left(\langle \partial_\theta \psi_\theta | \partial_\theta \psi_\theta \rangle - \left|\langle \partial_\theta \psi_\theta |\psi_\theta \rangle\right|^2\right) \\
    &= - \frac{1}{\theta^2\ln{\theta}}.
\end{align}
We observe that the quantum Fisher information is identical to the classical Fisher information, Eq.~\eqref{eq:CRB}, and thus any quantum bound of the coherent state, Eq.~\eqref{eq:coherent}, is identical to the CRB of the Poisson distribution, Eq.~\eqref{eq:p_poisson}.

\end{document}